\documentclass{sig-alternate}

\def\more-auths{%
\end{tabular}
\begin{tabular}{c}}

\begin{document}

\conferenceinfo{SIGIR'06,} {August 6--11, 2006, Seattle, Washington, USA.}

\CopyrightYear{2006}

\crdata{1-59593-369-7/06/0008} 

\title{ProbFuse: A Probabilistic Approach to Data Fusion}

\numberofauthors{3}

\author{
\alignauthor David Lillis \\
   \affaddr{School of Computer Science and Informatics} \\
   \affaddr{University College Dublin} \\
   \email{david.lillis@ucd.ie}
\alignauthor Fergus Toolan \\
   \affaddr{School of Computer Science and Informatics} \\
   \affaddr{University College Dublin} \\
   \email{fergus.toolan@ucd.ie}
\vskip1pc
\more-auths
\alignauthor Rem Collier \\
   \affaddr{School of Computer Science and Informatics} \\
   \affaddr{University College Dublin} \\
   \email{rem.collier@ucd.ie}
\alignauthor John Dunnion \\
   \affaddr{School of Computer Science and Informatics} \\
   \affaddr{University College Dublin} \\
   \email{john.dunnion@ucd.ie}
}


\maketitle
\begin{abstract}
Data fusion is the combination of the results of independent searches on a document collection into one single output result set. It has been shown in the past that this can greatly improve retrieval effectiveness over that of the individual results.

 This paper presents \emph{probFuse}, a probabilistic approach to data fusion. \emph{ProbFuse} assumes that the performance of the individual input systems on a number of training queries is indicative of their future performance. The fused result set is based on probabilities of relevance calculated during this training process. Retrieval experiments using data from the TREC ad hoc collection demonstrate that \emph{probFuse} achieves results superior to that of the popular CombMNZ fusion algorithm.

\end{abstract}




\section{Introduction} \label{introduction}

 In the past, many algorithms have been developed to address the Information Retrieval (IR) task of identifying which documents in a database are most relevant to a given topic or query. More recently, researchers have focussed on attempting to improve upon the performance of individual IR models by combining the outputs of a number of such models into a single result set \cite{lee97analyses} \cite{montague02condorcet} \cite{voorhees94collection}. 

The task of fusing result sets produced by using a number of IR models to query the same document collection has become known as \emph{data fusion} \cite{aslam00bayes}. This is different to \emph{collection fusion} \cite{voorhees94collection}, which involves the fusion of result sets that have been produced from querying distinct document collections that have little or no overlap.

This paper is organised as follows: In section \ref{probdesc} we describe the problem that fusion is intended to solve. Section \ref{background} outlines previous work that has been undertaken by other researchers in this field. In section \ref{probFuse} we describe \emph{probFuse}, a novel probabilistic algorithm for data fusion. Section \ref{experiments} describes our experiments to evaluate the performance of the \emph{probFuse} algorithm on inputs taken from two Text REtrieval Conferences (TREC). We also compare this with the performance of the popular CombMNZ approach \cite{fox94combination}. Finally, our conclusions and future work are outlined in section \ref{conclusions}.

\section{Problem Description} \label{probdesc}

Of the numerous approaches to IR that have been proposed, none has been shown to achieve superior performance to all others in all situations. This may be as a result of difference in policies regarding query or document preprocessing, the algorithms used and representations of documents and queries. Individual IR systems have been shown to retrieve different documents in response to the same queries when operating on the same document collection \cite{dasgupta83study}. This has been observed even where the overall retrieval effectiveness of these different systems has been similar \cite{harman93overview}.

Retrieval performance has been shown to be improved by fusing the result sets produced by a number of different IR systems into a single result set. A number of different approaches to data fusion are outlined in section \ref{background}.

Vogt and Cottrell \cite{vogt99linear} identify three ``effects'', any of which can be leveraged by a fusion technique. In some cases, a number of input result sets agree on the relevance of particular documents. Fusion techniques that take this agreement into account when compiling the fused result set will perform well in such circumstances. This is described as the ``Chorus Effect''. Experiments carried out by Lee \cite{lee97analyses} have shown that this is a very significant effect for data fusion tasks. 

The exploitation of the Chorus Effect is the principal difference between the data fusion and collection fusion tasks. For data fusion, each individual IR technique is searching an identical document collection. This means that whenever a document is contained in multiple result sets, this can be presumed to infer relevance. However, when the document collections being searched are disjoint (collection fusion), this situation clearly cannot arise. In situations where the collections are partially overlapping, the presence of a document in multiple result sets cannot be used as an indication of greater relevance than a document that only appears in one. A document may only appear in a single result set because either the other IR models did not consider it to be relevant to the given query or it was not contained in the other document collections. Thus, in order for a fusion technique to make use of the Chorus Effect, it must be known that the document collections that are being queried by the different inputs have a very high degree of overlap.

They also describe the ``Skimming Effect''. Multiple result sets are more likely to result in higher recall (i.e. the fraction of relevant documents that have been retrieved) than a single one. A fusion technique can take advantage of this by ``skimming'' the top documents from each result set, as that is where the relevant documents are most likely to occur.

The ``Dark Horse Effect'' is an apparent contradiction to the Chorus Effect. Here, fusion effectiveness can be improved by identifying one input result set whose quality is of a substantially different level to the others. This may be due to either unusually high or unusually low numbers of relevant documents being returned. The apparent contradiction arises because while the Chorus Effect argues that fusion techniques should take as many of the input result sets as possible into account, the Dark Horse Effect argues in favour of identifying a single input result set

\section{Background Research} \label{background}
There are two principal categories of fusion techniques. Some algorithms make use of the score assigned to each document in each input result set to calculate a final score for that document in the fused result set. Because these raw scores are not always directly comparable (e.g. one input result set might assign scores in a range of 0-100 while another uses 0-1), score-based techniques frequently make use of a score normalisation phase before fusion takes place. This typically involves the mapping of all scores to a common range. Others make use of the rank each document occupies in each of the inputs, as the scores are not always available. 

A Linear Combination model has been used in a number of studies \cite{bartell94automatic} \cite{callan95searching} \cite{powell00impact} \cite{vogt99linear}. Under this model, a weight is calculated for each input model. In order to calculate a ranking score for a document, the score it is assigned by each input model is multiplied by that model's weight. These are then summed to get the final ranking score for the fused result set. A variation on the Linear Combination model using normalised scores was used in \cite{larkey00collection} and \cite{si02distributed}.

A number of fusion techniques based on normalised scores were proposed by Fox and Shaw \cite{fox94combination}. Of these, CombSum and CombMNZ were shown to achieve the best performance and have been used in subsequent research. Under CombSum, a document's score is calculated by adding the normalised scores returned by the individual input models. Its CombMNZ score is found by multiplying the CombSum score by the number of non-zero relevance scores that it was assigned. In particular, Lee \cite{lee97analyses} achieved positive results using CombMNZ on the TREC-3 data set. These techniques have also been used in real-world systems: The \emph{MetaCrawler} \cite{selberg97metacrawler} and \emph{SavvySearch} \cite{howe97savvysearch} meta search engines both use CombSum to fuse results.

Research by Manmatha et al. \cite{manmatha01modeling} demonstrated that the scores given to documents by an IR system can be modelled using a normal distribution for relevant documents and an exponential distribution for nonrelevant documents. This was possible even when relevance judgments were not available for the queries in question. Using Bayes' Rule, it was then possible to calculate the probability of relevance, given the score. When performing fusion, these probabilities were averaged, producing performance approaching that of CombMNZ.

Perhaps the simplest rank-based fusion technique is \emph{interleaving} \cite{voorhees94collection}. Under this system, the fused result set is constructed by firstly taking the top-ranked document from each input result set, followed by the second-ranked documents and so on. This approach operates on the assumption that each of the inputs are of similar effectiveness, and has been shown empirically to fall short of its goal of outperforming its inputs \cite{voorhees95learning}. Voorhees et al. \cite{voorhees95learning} proposed two variations on simple interleaving, in which training data was used to weight the input models according to past performance. At the interleaving stage, different quantities of documents were taken from each result set, depending on these weights, rather than taking equal amounts from each.

Lee \cite{lee97analyses} proposed a rank-based variation on CombMNZ, in which a function of each document's rank in each input result set was used as an alternative for normalised scores.

Aslam and Montague compared the fusion process to a democratic election in which there are few voters (the input models) and many candidates. They achieved positive results by implementing adapted implementations of two algorithms designed for that situation. \emph{Borda-fuse} \cite{aslam01models} awards a score to each document depending on its position in each input result set, with its final score being the sum of these. \emph{Condorcet-fuse} \cite{montague01relevance} ranks documents based on a pairwise comparison of each. A document is ranked above another if it appears above it in more input result sets.

Other techniques have been proposed that make use of the actual textual content of the documents returned \cite{craswell99merging} \cite{lawrence98inquirus}. Others rely on the individual input models providing metadata about the returned documents, other than simply a ranked list with relevance scores \cite{gravano97starts}.

\section{Probabilistic Fusion} \label{probFuse}
In this section, we describe \emph{probFuse}, a probabilistic approach to data fusion. \emph{ProbFuse} ranks documents based on their probability of relevance to the given query. This probability is calculated during a training phase, and depends on which input system returned the document amongst its results and the position in the result set in which the document was returned.

The inputs to the fusion process are a number of collections of result sets that are produced by different IR models running the same queries on the same document collection. In order to run \emph{probFuse}, we first build a set of probabilities for each input set. These probabilities are calculated by analysing the performance of each individual model on a number of training queries.

Rather than using the exact position a document occupies in each result set, the input result sets are divided into $x$ segments. For each segment, the probability that a document being returned in this segment is relevant to a given query is calculated. This probability is averaged over $t\%$ of the total queries that are available. 

In a training set of $Q$ queries, $P(d_k|m)$, the probability that a document $d$ returned in segment $k$ is relevant, given that it has been returned by retrieval model $m$, is given by:

\begin{equation} \label{probFuseAll}
P(d_k|m) = \frac{\sum_{q=1}^Q\frac{|R_{k,q}|}{|k|}}{Q}
\end{equation}

where $|R_{k,q}|$ is the number of documents in segment $k$ that are judged to be relevant to query $q$, and $|k|$ is the total number of documents in segment $k$.

In the past, it has been demonstrated that \emph{probFuse} achiev\-es significantly better results than CombMNZ when applied to small document collections \cite{lillis05probability}. For these collections, full relevance judgments are available, so the relevance of every document is known during the training phase. For larger collections, however, this is not the case, as the relevance judgments are incomplete (i.e. for some documents, it is unknown whether they are relevant or nonrelevant to the given queries). For this reason, we also use a slight variation of the probability calculation. This allows us to observe the effects, if any, of different methods of dealing with unjudged documents.

Equation \ref{probFuseAll} takes all the documents in a segment into account, assuming unjudged documents to be nonrelevant. Our modified probability calculation ignores unjudged documents and thus only takes into account documents that have been judged to be either relevant or nonrelevant. In this case, the probability $P(d_k|m)$ is given by

\begin{equation} \label{probFuseJudged}
P(d_k|m) = \frac{\sum_{q=1}^Q\frac{|R_{k,q}|}{|R_{k,q}|+|N_{k,q}|}}{Q}
\end{equation}

where $|R_{k,q}|$ is the number of documents in segment $k$ that are judged to be relevant to query $q$, and $|N_{k,q}|$ is the number of documents in segment $k$ that are judged to be nonrelevant to query $q$.

We refer to \emph{probFuse} runs using the probability calculation in equation \ref{probFuseAll} as \emph{probFuseAll} and those using equation \ref{probFuseJudged} as \emph{probFuseJudged}. From these equations, it can be seen that for document collections with complete relevance judgments, the probabilities calculated by \emph{probFuseAll} and \emph{probFuseJudged} will be identical.

After the training phase is complete and a set of probabilities for each input model has been built, we can then use this to construct a fused result set for subsequent queries. For these, the ranking score $S_d$ for each document $d$ is given by

\begin{equation}
S_d = \sum_{m=1}^M\frac{P(d_k|m)}{k}
\end{equation}

where $M$ is the number of retrieval models being used, $P(d_k|m)$ is the probability of relevance for a document $d_k$ that has been returned in segment $k$ in retrieval model $m$, and $k$ is the segment that $d$ appears in (1 for the first segment, 2 for the second, etc.).  For any input model that does not return document $d$ in its result set at all, $P(d_k|m)$ is considered to be zero, in order to ensure that documents do not receive any boost to their ranking scores from models that do not return them as being judged relevant.

This approach to data fusion attempts to make use of the three effects described in section \ref{probdesc} above. By using the sum of the probabilities, we attach more significance to documents that have been returned by multiple input models, thus exploiting the Chorus Effect. The division by the segment number $k$ gives a greater weight to documents that appear early in each of the individual result sets, making use of the Skimming Effect. Finally, because the probabilities are calculated based on the actual past performance of each input model, we attach greater importance to input models that are more likely to return relevant documents in particular segments (Dark Horse Effect).

\section{Experiments and Evaluation} \label{experiments}

In this section, we describe the experiments that were performed to evaluate the effectiveness of \emph{probFuse}. The \emph{probFuse} algorithm was applied to a number of different combinations of input result sets and the resulting fused result was compared to that of the popular CombMNZ algorithm. CombMNZ is easily implemented and has been shown to perform well on data fusion tasks \cite{lee97analyses}. This has made it an attractive choice when choosing a baseline technique to compare with. As such, it has become the standard algorithm to use \cite{beitzel04effective} \cite{montague02condorcet}. 

In order to run CombMNZ, two steps must be performed. Firstly, the scores attributed to each document by each input must be normalised, so that they lie in a common range. A number of different normalisation strategies have been proposed. We have chosen the one used by Lee \cite{lee97analyses}, as it is the one most commonly used for comparison and has been described as ``Standard Normalisation'' \cite{montague01relevance}. Lee's implementation of CombMNZ calculates normalised scores using

\begin{equation}
normalised\_sim = \frac{unnormalised\_sim - min\_sim}{max\_sim - min\_sim}
\end{equation}

where $max\_sim$ and $min\_sim$ are the maximum and minimum scores that are actually seen in the input result set. Once the scores have been normalised, $CombMNZ_d$, the CombMNZ ranking score for any document $d$ is given by

\begin{equation}
CombMNZ_d = \sum_{s=1}^S N_{s,d} \times |N_d > 0|
\end{equation}

where $S$ is the number of result sets to be fused, $N_{s,d}$ is the normalised score of document $d$ in result set $s$ and $|N_d > 0|$ is the number of non-zero normalised scores given to $d$ by any result set.

\subsection{Experimental Setup}

As our inputs, we used data from the ad hoc retrieval track of the TREC-3 and TREC-5 conferences. This data consists of the topfile (a collection of result sets) produced by each of the groups that participated in those conferences. Each topfile contains result sets for 50 topics (queries): TREC-3 uses TREC topics 151-200 and TREC-5 uses topics 251-300. 40 topfiles are available for TREC-3, while 31 are available for TREC-5. Only the topfiles in Category A were considered for TREC-5, as Category B participants operated on only a subset of the data. 

\begin{table}
\caption{Inputs to TREC-3 experimental runs}
\begin{center}
\begin{tabular}{l l l l l} \label{3inputs}
first & second & third & fourth & fifth \\
\hline
acqnt1 & clartm & brkly7 & assctv1 & assctv2 \\
citri1 & crnlla & clarta & erima1 & nyuir1 \\
crnlea & dortd2 & dortd1 & lsia0mf & rutfua1 \\
padre2 & eth002 & eth001 & lsia0mw2 & rutfua2 \\
xerox3 & nyuir2 & inq101 & virtu1 & siems1 \\
xerox4 & padre1 & pircs1 & vtc2s2 & westp1 \\
\end{tabular}
\end{center}
\end{table}

\begin{table}
\caption{Inputs to TREC-5 experimental runs}
\begin{center}
\begin{tabular}{l l l l l} \label{5inputs}
first & second & third & fourth & fifth \\
\hline
brkly18 & anu5man4 & anu5aut2 & DCU962 & anu5aut1 \\
DCU963 & CLCLUS & city96a1 & genrl1 & colm4 \\
ETHal1 & erliA1 & CLTHES & ibmge1 & Cor5A2cr \\
KUSG3 & genrl3 & ETHas1 & ibms96a & LNmFull1 \\
vtwnA1 & ibms96b & genrl4 & KUSG2 & LNmFull2 \\
vtwnB1 & uwgcx0 & ibmgd2 & mds003 & pircsAAL \\
\end{tabular}
\end{center}
\end{table}

For each of these two data sets, we performed 5 experimental runs. Each experimental run firstly involved choosing six random topfiles to use as inputs. In order to eliminate the results being skewed by the ordering of the topics, we produced five random orderings for the topics and performed data fusion using both \emph{probFuse} and CombMNZ over each of these. The performance evaluation values for each run are the average for each of those five random orderings. No input topfile was used in multiple runs. The inputs used for each of the five runs for TREC-3 and TREC-5 are shown in Table \ref{3inputs} and Table \ref{5inputs} respectively. The inclusion of this list of inputs is intended to aid the reproduction of our experiments.

\begin{table*}
\caption{TREC-3 Average MAP scores for various training set sizes}
\begin{center}
\begin{tabular}{l|l|l l|l l} \label{3map}
Training t\% & CombMNZ & \emph{probFuse} & Difference & \emph{probFuse} & Difference\\
 & & \emph{All} & v. MNZ & \emph{Judged} & v. MNZ\\
\hline
10\% & 0.29593 & 0.33885 & +14.50\% & 0.33988 & +14.85\% \\	
20\% & 0.29738 & 0.34146 & +14.82\% & 0.34312 & +15.38\% \\
30\% & 0.30134 & \textbf{0.34628} & +14.91\% & \textbf{0.34830} & +15.58\% \\
40\% & 0.29753 & 0.34307 & +15.31\% & 0.34517 & +16.01\% \\
50\% & 0.29557 & 0.34230 & \textbf{+15.81}\% & 0.34445 & \textbf{+16.54}\%
\end{tabular} 
\end{center} 
\end{table*}

\begin{table*}
\caption{TREC-5 Average MAP scores for various training set sizes}
\begin{center}
\begin{tabular}{l|l|l l|l l} \label{5map}
Training t\% & CombMNZ & \emph{probFuse} & Difference & \emph{probFuse} & Difference\\
 & & \emph{All} & v. MNZ & \emph{Judged} & v. MNZ \\
\hline
10\% & 0.17842 & 0.26011 & +45.79\% & 0.25987 & +45.65\% \\	
20\% & 0.17604 & 0.25959 & \textbf{+47.46}\% & 0.26020 & +47.81\% \\
30\% & 0.17528 & 0.25842 & +47.43\% & 0.25937 & \textbf{+47.98\%} \\
40\% & 0.17720 & 0.25959 & +46.50\% & 0.26056 & +47.04\% \\
50\% & 0.17712 & \textbf{0.26061} & +47.14\% & \textbf{0.26175} & +47.78\%
\end{tabular} 
\end{center} 
\end{table*}

Each run was performed for a variety of training set sizes defined as a percentage of the number of available queries. In the case of the TREC-3 and TREC-5 data, the number of available queries is always 50. The number of segments into which each result set was divided was also varied. We used training set sizes, $t$ such that $t \in$\{10, 20, 30, 40, 50\} and numbers of segments, $x$ such that $x \in$\{2, 4, 6, 8, 10, 15, 20, 25, 30, 40, 50, 100, 150, 200, 250, 300, 400, 500\}.

In order to ensure comparability, CombMNZ was only applied to those queries used in the fusion phase of \emph{probFuse}, with the training queries being ignored. This will cause the evaluation results for CombMNZ to vary as the training set size changes, since the number of remaining queries on which fusion is performed also changes.

The goal of these experiments is to empirically determine the combination of training set size and number of segments (denoted by $x$) that achieves the greatest retrieval performance, both in terms of the evaluation scores it receives and its performance in relation to CombMNZ. We approach this by firstly identifying a training set size that results in high performance for both the TREC-3 and TREC-5 input sets. This is discussed in section \ref{training_set_size}. Once this is done, we examine the performance of both variations of \emph{probFuse} for different values of $x$ to find an $x$-value that performs well on both input sets when averaged over the five runs. This is done in section \ref{x-value}. 

The evaluation of the fused output result sets was performed by \emph{trec\_eval}, which is the evaluation software used for the TREC conferences \cite{voorhees00overview}. We use two evaluation measures in our experiments. Firstly, we use Mean Average Precision (MAP) to find our training set size and $x$-value. MAP is the mean of the precision scores obtained after each relevant document is retrieved, using zero as the precision for relevant documents that are not retrieved \cite{buckley04incomplete}. Documents for which a relevance judgment is not available are considered to be nonrelevant. 

After identifying this training set size and $x$-value, we then examine each of the five experimental runs to make a comparison with the CombMNZ algorithm in a more detailed manner. At this stage, in addition to MAP we also make use of the bpref evaluation measure. Bpref only takes judged documents into account and is inversely related to the fraction of judged nonrelevant documents that are retrieved before relevant documents \cite{buckley04incomplete}. The analysis of these evaluation results is contained in section \ref{runs}.

\subsection{Training Set Size} \label{training_set_size}

Initially, the MAP measure was used to identify which training set sizes resulted in the best performance. Table~\ref{3map} and Table~\ref{5map} show the average MAP achieved when using each training set size, along with its improvement over the corresponding figure for CombMNZ. The MAP score included in those figures is the average MAP score for all values of $x$ (the number of segments) at that training set size over all five runs. The average MAP scores for CombMNZ vary with training set size, despite CombMNZ not making use of any training phase. In each of those tables, the highest MAP score and the greatest improvement over CombMNZ for each of the \emph{probFuse} variants are marked in bold type.

\begin{figure}[h]
\begin{center}
\includegraphics[height=200pt,width=250pt]{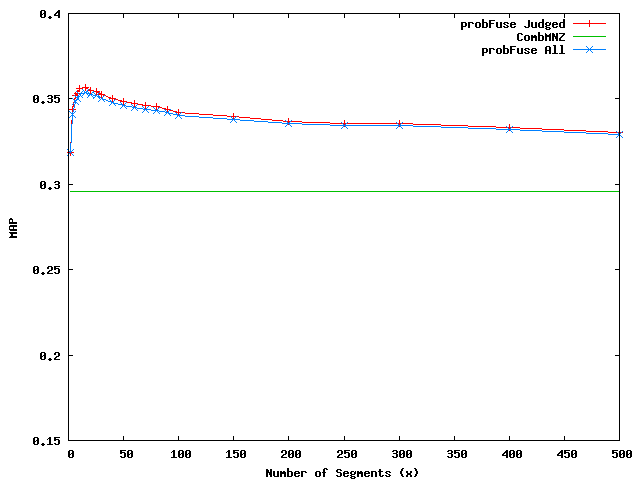}

\caption{TREC-3 MAP scores for $t=50\%$ (average over 5 runs)}\label{3average}
\end{center}
\end{figure}

\begin{figure}[!h]
\begin{center}
\includegraphics[height=200pt,width=250pt]{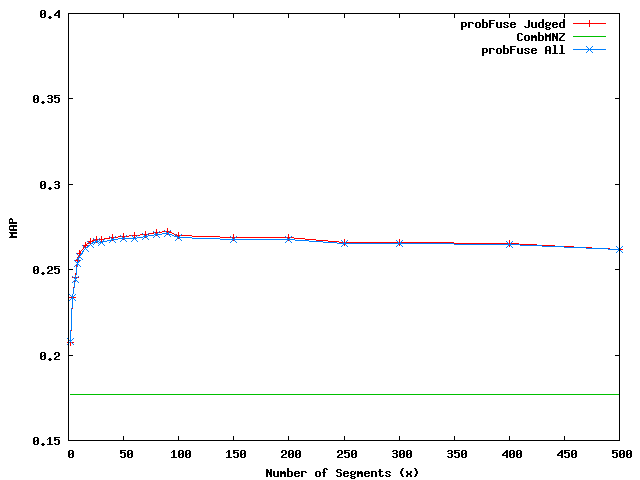}

\caption{TREC-5 MAP scores for $t=50\%$ (average over 5 runs)}\label{5average}
\end{center}
\end{figure}

From Table~\ref{3map}, we can see that the highest average MAP score by both \emph{probFuse} variations on the TREC-3 inputs was achieved using a training set size of 30\%. The biggest percentage increase over CombMNZ was when using a training set size of 50\%. For the TREC-5 inputs (seen in Table~\ref{5map}), we note that the highest average MAP score was when using a training set size of 50\%. The greatest percentage improvement over CombMNZ was achieved for training set sizes of 20\% and 30\% for \emph{probFuseAll} and \emph{probFuseJudged} respectively.

Both variations of \emph{probFuse} achieve higher average MAP scores than CombMNZ for every training set size. This is the case for both the TREC-3 and TREC-5 input sets.

For the purposes of Section \ref{x-value}, we have chosen to use a training set size of 50\%. At that level, \emph{probFuseAll} and \emph{probFuseJudged} both achieve either their highest MAP score or their highest increase over CombMNZ for both sets of inputs. This would not be the case had we chosen a training set size of 30\%, as \emph{probFuseAll} achieves its biggest improvement over CombMNZ at a training set size of 20\%. At 50\%, both \emph{probFuse} variations achieve their highest average MAP scores on the TREC-5 inputs. For the TREC-3 inputs, both variations achieve their highest improvement over CombMNZ, and their average MAP scores are within 2\% of the highest they achieve at any level. 

\subsection{Number of Segments} \label{x-value}

Figure~\ref{3average} and Figure~\ref{5average} show the MAP scores for each number of segments for a training set size of 50\% for the TREC-3 and TREC-5 inputs respectively. Each of these MAP scores is the average of the MAP scores achieved in each of the five runs.

It is interesting to note that \emph{probFuseJudged} and \emph{probFuseAll} both show near-identical results for both input sets. In each graph, performance is at its worst for a value of $x=2$. This is to be expected, as in that situation, each result set is being divided into only two segments, so the probability of relevance being assigned to each document is based on whether it appears in the first or second half of the result set, which is too coarse a measure. Initially, as $x$ increases, the average MAP score improves. A gradual decline is then seen for higher values of $x$.

The principal difference in the trend in the two graphs is in the point at which the average MAP score reaches its peak. This peak is reached for a much lower $x$-value on the TREC-3 inputs, where $x=15$. For TREC-5, the MAP score continues improving gradually until the point where $x=90$. Thereafter, both show a downward trend as $x$ increases.

An $x$-value of 25 yields the best MAP score, on average, over both input sets. For that reason, this is the $x$-value we have chosen to use when analysing the individual runs in section~\ref{runs}.

\subsection{Analysis of Individual Runs} \label{runs}

Having identified a high-performing training set size (50\%) and number of segments to divide each result set into (25), we can examine the five individual experimental runs in more detail.

\begin{table*}
\caption{TREC-3 performance of five individual runs for $t=50\%$ and $x=25$}
\begin{center}
\begin{tabular}{l|l l|l l|l l} \label{3runs}
        & CombMNZ & & \emph{probFuseAll} & & \emph{probFuseJudged} & \\
        & MAP     & bpref   & MAP     & bpref   & MAP & bpref \\
\hline
first   & 0.16726 & 0.23960 & 0.30988 (+85.27\%) & 0.31458 (+31.29\%) & 0.31144 (+86.20\%) & 0.31628 (+32.00\%) \\
second  & 0.28752 & 0.33434 & 0.34100 (+18.60\%) & 0.33118 (-0.95\%) & 0.34402 (+19.65\%) & 0.33356 (-0.23\%) \\
third   & 0.43344 & 0.41222 & 0.41348 (-4.61\%) & 0.39222 (-4.85\%) & 0.41620 (-3.98\%) & 0.39416 (-4.38\%) \\
fourth  & 0.23416 & 0.31048 & 0.30374 (+29.71\%) & 0.30314 (-2.36\%) & 0.30766 (+31.39\%) & 0.30528 (-1.67\%) \\
fifth   & 0.35548 & 0.39616 & 0.39108 (+10.01\%) & 0.38006 (-4.06\%) & 0.39294 (+10.54\%) & 0.38308 (-3.30\%) \\
\hline
Average & 0.29557 & 0.31707 & 0.35184 (+19.04\%) & 0.34804 (+9.77\%) & 0.35445 (+19.92\%) & 0.35046 (+10.53\%) \\
\end{tabular}
\end{center}
\end{table*}

\begin{figure}[!h]
\begin{center}
\includegraphics[scale=0.5]{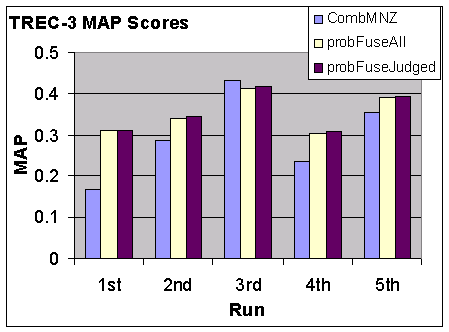}

\caption{TREC-3 MAP scores for $t=50\%$ and $x=25$}\label{3runsmap}
\end{center}

\begin{center}
\includegraphics[scale=0.5]{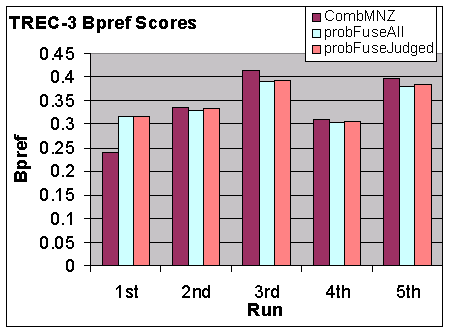}

\caption{TREC-3 bpref scores for $t=50\%$ and $x=25$}\label{3runsbpref}
\end{center}
\end{figure}

Table~\ref{3runs} shows the results of the five individual runs on the TREC-3 input set. In that table, figures in parentheses represent the percentage difference to the corresponding score for CombMNZ. Figure~\ref{3runsmap} shows the MAP scores for CombMNZ, \emph{probFuseAll} and \emph{probFuseJudged} for each run on the TREC-3 data, and figure~\ref{3runsbpref} shows the bpref scores for TREC-3.

Both variants of \emph{probFuse} achieved a higher MAP score than CombMNZ for all runs except for ``third''. On that run, CombMNZ scored slightly higher. It is important to highlight that both \emph{probFuse} variations actually achieve their highest MAP scores for that run. The MAP score for Comb-MNZ is also the highest it achieves for any run. The lower MAP score achieved by \emph{probFuse} on the ``third'' run can therefore be attributed to an unusually high MAP score being achieved by CombMNZ on that run, rather than \emph{probFuse} underperforming.

Using the bpref measure, the performance of \emph{probFuse} is slightly below that of CombMNZ for four of the five runs. For this reason, it cannot be said that \emph{probFuse} outperforms CombMNZ on the TREC-3 inputs under bpref. However, neither \emph{probFuse} variant drops below 95\% of CombMNZ's bpref score on any run, and for the ``first'' run, \emph{probFuse} achieves a vastly superior result, leading to a better average performance.

Under both the MAP and bpref measures, \emph{probFuseJudged} achieves higher performance than \emph{probFuseAll} on each of the five experimental runs. However, this increase only exceeds 1\% for the MAP score on the ``fourth'' run, and never exceeds 2\%.

\begin{table*}
\caption{TREC-5 performance of five individual runs for $t=50\%$ and $x=25$}
\begin{center}
\begin{tabular}{l|l l|l l|l l} \label{5runs}
        & CombMNZ & & \emph{probFuseAll} & & \emph{probFuseJudged} & \\
        & MAP     & bpref & MAP & bpref & MAP & bpref \\
\hline
first   & 0.25144 & 0.26406 & 0.27378 (+8.88\%) & 0.26814 (+1.55\%) & 0.26264  (+4.45\%) & 0.26878 (+1.79\%) \\
second  & 0.22480 & 0.26896 & 0.35560 (+58.19\%) & 0.33918 (+26.11\%) & 0.35844 (+59.45\%) & 0.34140 (+26.93\%) \\
third   & 0.12306 & 0.19232 & 0.26838 (+118.09\%) & 0.24744 (+28.66\%) & 0.27050 (+119.81\%) & 0.24920 (+29.58\%) \\
fourth  & 0.12626 & 0.14520 & 0.15546 (+23.13\%)& 0.15734 (+8.36\%) & 0.15602 (+23.57\%) & 0.15746 (+8.44\%) \\
fifth   & 0.16004 & 0.21790 & 0.27842 (+73.97\%)& 0.26474 (+21.50\%) & 0.27922 (+74.47\%) & 0.26498 (+21.61\%) \\
\hline
Average & 0.17712 & 0.19740 & 0.26633 (+50.37\%) & 0.25537 (+29.37\%) & 0.26787 (+51.24\%) & 0.26212 (+32.79\%)
\end{tabular}
\end{center}
\end{table*}

\begin{figure}[!h]
\begin{center}
\includegraphics[scale=0.5]{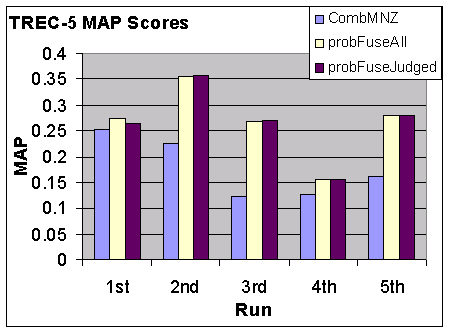}

\caption{TREC-5 MAP scores for $t=50\%$ and $x=25$}\label{5runsmap}
\end{center}

\begin{center}
\includegraphics[scale=0.5]{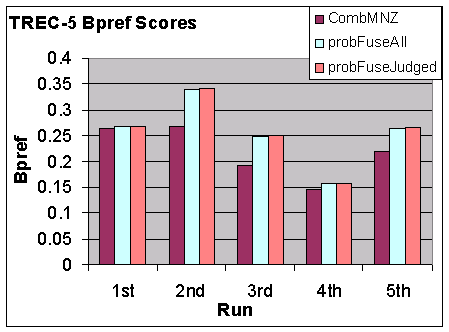}

\caption{TREC-5 bpref scores for $t=50\%$ and $x=25$}\label{5runsbpref}
\end{center}
\end{figure}

Table~\ref{5runs} shows a similar table detailing the five individual experimental runs on the TREC-5 input set. Figure~\ref{5runsmap} shows the MAP scores for CombMNZ, \emph{probFuseAll} and \emph{probFuseJudged} for each run on the TREC-5 data, and figure~\ref{5runsbpref} shows the bpref scores for each fusion technique.

Here, \emph{probFuse} outperforms CombMNZ on each of the runs using both evaluation measures. In particular, runs ``second'', ``third'' and ``fifth'' show large performance gains for both \emph{probFuse} variations over CombMNZ for both the MAP and bpref measures. 

As with TREC-3, \emph{probFuseJudged} performs better than \emph{probFuseAll}, although once again the degree of improvement is less than 1\% in almost all cases. The exception to this is the MAP score for the ``first'' run, which is the only case where \emph{probFuseAll} performs better than \emph{probFuseJudged}. This is particularly interesting in the case of the bpref scores, as bpref only takes judged documents into account, ignoring documents for which relevance judgments are not available. For this level of incompleteness, the performance of \emph{probFuseAll} is similar to that of \emph{probFuseJudged}. It is left to future work to determine if this remains the case as the available relevance judgments become more incomplete.

The performance on the TREC-3 inputs is superior to CombMNZ when evaluated using the MAP measure, with the bpref scores falling only slightly below those of CombMNZ in some cases. For the TREC-5 inputs, \emph{probFuse} has shown significant improvement over CombMNZ when evaluated by both MAP and bpref.

\section{Conclusions and Future Work} \label{conclusions}

In this paper, we have described \emph{probFuse}, a data fusion algorithm that relies on the probability of relevance to calculate a ranking score for documents in a fused result set. These probabilities are calculated based on the position of relevant documents in result sets returned in response to a number of training queries.

In experiments using data from the ad hoc track of the TREC-3 and TREC-5 conferences, two variations of \emph{probFuse} were shown to significantly outperform the popular CombMNZ algorithm over a number of different combinations of inputs. \emph{ProbFuseAll} achieved a MAP score that was, on average, 19\% higher than CombMNZ on the TREC-3 inputs and 50\% higher on TREC-5. Similarly, the average bpref score was 10\% higher than CombMNZ on TREC-3 inputs and 29\% higher on TREC-5. 

These results follow on from experiments on small document collections for which complete relevance judgments are available \cite{lillis05probability}. Due to the incomplete nature of the relevance judgments for TREC-3 and TREC-5, we also tested a variation of \emph{probFuse}, called \emph{probFuseJudged}, that only takes judged documents into account when calculating its probabilities. \emph{ProbFuseJudged} achieved an increase of 11\% over CombMNZ on the TREC-3 inputs and an increase of 31\% on TREC-5. The MAP scores it achieved were 20\% higher than CombMNZ on TREC-3 and 51\% higher on the TREC-5 data. 

It is interesting to note that \emph{probFuseJudged} only achieved marginal performance gains over \emph{probFuseAll}, even using the bpref evaluation measure, which only takes judged documents into account. 

Our future work will apply \emph{probFuse} to larger collections with greater levels of incompleteness in their available relevance judgments (e.g. the Web track collections of later TREC conferences). We intend to investigate whether the performance of \emph{probFuseJudged} and \emph{probFuseAll} diverges as the level of incompleteness increases. In addition, the relevance judgments for some collections differentiate between different degrees of relevance (e.g. for the WT10G collection, documents can be judged nonrelevant, relevant and highly relevant). We also intend to investigate whether adjustments to our probability calculations that take this information into account will be beneficial.

Another potential research direction is to investigate the possibilities of applying \emph{probFuse} to a document collection without the necessity of using training data from that collection. This could potentially involve the use of the techniques outlined by Manmatha et al. \cite{manmatha01modeling} to estimate the probability of relevance, or alternatively training \emph{probFuse} on one document collection in order to apply it to another.

\bibliographystyle{abbrv}
\bibliography{master} 

\end{document}